\documentclass[journal]{IEEEtran}
\ifCLASSINFOpdf
\else
\fi
\usepackage{circuitikz}
\usepackage{tikz}
\usetikzlibrary{positioning}
\begin{document}
\title{A New Algorithm to Compare the Magnitude of Two
RNS Numbers}
\author{Parham ~Ghayour~\IEEEmembership{}
\thanks{Parham Ghayour is with the Department
of Electrical and Computer Engineering, University Of Southern California, CA, 90017, USA e-mail : ghayour@usc.edu
.}}
\maketitle
\begin{abstract}
Comparison of two numbers in RNS systems is a challenging task. In this paper, a new algorithm to compare the magnitude of two RNS numbers, using a clustering method has been proposed. In the clustering process, each inputted number is assigned to a cluster. To compare the magnitude of two numbers, first the clusters of these numbers and their differences are obtained. Then by comparing these clusters, the relative magnitude of two numbers is determined. All of these processes are performed in RNS system without converting numbers to the binary system.
\end{abstract}
\IEEEpeerreviewmaketitle
\section{Introduction}
Since the advent of the residue number system, it has
attracted considerable attention, mostly because it has carry
free arithmetic operations. One of the difficulties of the
residue number system is the determination of the relative
magnitude of two numbers. There have been many techniques
that are proposed as a solution but all of them have some
limitations.
The traditional techniques for magnitude comparison in RNS
use the Chinese Remainder Theorem  or the Mixed
Radix Conversion to convert the numbers from the
residues to a positional code [3], [6]. However both these
techniques are inefficient, because CRT requires modulo $M$
operations (where $M$ is the range of the number system) and
MRC is a slow sequential method. Another technique for
comparing the magnitude of numbers in residue
representation, originally proposed by Akushskii, Burcev and
Park, uses the concept of “core function”.The authors
proposed and applied a descendent and lift scheme to determine the “critical core” values. An improved version of
this technique has been proposed several years later,
avoiding the iterative procedure at the cost of introducing a
redundant modulus [2]. Another proposed approach for
magnitude number comparison in RNS is based on the
“diagonal function”, defined as the sum of suitable quotients
for estimating its’ magnitude order [1] . A new algorithm
based on the New Chinese Reminder Theorems has also been
proposed in [4] to compare the magnitude of numbers in RNS
[5]. By applying the new CRT II, it reduces the CRT modulo
operation size to $\sqrt{M}$. Recently, a new method is proposed for magnitude comparison that is based on two pairs of conjugate moduli. All of these techniques have some limitations in choosing modulo set.
In this paper a new technique for magnitude comparison in RNS numbers is proposed. In the proposed method, the modulo set of RNS system is considered to be \{ t,p,q \}, while t,p and q are mutually relatively prime numbers. First, a cluster is assigned to each number. Then, by comparing these clusters, the relative magnitude of two numbers is determined. All of these processes are performed in RNS system without converting numbers to the binary system. The rest of paper is organized as follows. In section II, the clustering process, the core process of the proposed algorithm is presented. The main algorithm to compare two RNS numbers using clustering method is introduced in section III. In Section IV the Boolean circuit of the algorithm  for the modulo set  $(2,3,5)$ has been proposed and section V concludes the paper.

\section{Clustering Process}

The technique represented in the clustering process
determines that each number $(r_1,r_2,r_3)$ of modulo set $(p_1,p_2,p_3)$ is in which cluster . 
All numbers in the range of this RNS system, where $ M=p_1 \times p_2 \times p_3 $, is divided into $p_1$ clusters.
Cluster$[1]$ consists of numbers between zero to $(M/p_1)-1$ .
Cluster$[2]$ consists of numbers between $(M/p_1)$ to         $(2\times M/p_1)-1$ and therefore Cluster$[p_1]$ consists of numbers between $((p_1-1) \times M/p_1)$ to $(p_1 \times M/p_1)-1$.\par
Each one of the clusters could be categorized into $p_2$ groups
$(0, p_2-1)$, because the second residue $r_2$ of each number in each cluster vary from $(0,p_2-1)$. This is the list of the numbers of the cluster m in order of magnitude that have $r_2=r$, and $r \in ( 0,p_2-1 )$. We call each group by its’ $r_2$  component, table $(I)$. That is, group$(r)$ consists of numbers which second component $r_2$ is equal to r, $r_2=r$.
The other parameters in table $(I)$ are defined as follows. In
these definitions, a.mod.b is the residue of dividing a to b.\\\\
$ D(m) = (m-1) \times p_2 \times p_3          $\\
$ A(j,m) = (D(m) +(j \times p_2 + r)).$ mod$. p_1  $\\
$ B(j,m) = (D(m) +(j \times p_2 + r)).$ mod$. p_2 = r $\\
$ C(j,m) = (D(m)+(j \times p_2+r)).$ mod$. p_3 = ( j \times p_2 + r).$ mod$. p_3  $\\\\
Note:\\\\
1- $j \in ( 0\ to \ p_3-1 )$\\
2- $m \in (1 \ to   \  p_1) $\\
3- $B(j, m) = r $\\
4- $C(j, m) $ is independent of the value of m.\\  
In other words all of the numbers in table $(I)$ , $N(j,m)$, have $r_2$ or $B(j,m)$ equal to r. $A(j,m),B(j,m)$ and $C(j,m)$ are the residues of division of $N(j,m)$ to $p_1$, $p_2$ and $p_3$.\\
Now, if we calculate for each group the residues of the
division of each one of the fourth column residues $C(j,m)$ for
$ j=(0 \ to \ p_3-1) \ to \ p_2$, it can have $p_2$ different answers. We call each one of these answers a subgroup.\\\\
Note:\\ the equation below is for $i \in (0, p_2-1)$. \\\\
$ S(r,i)=r-(i \times (p_3.$mod$.p_2)).$mod$.p_2 $\\\\
If we combine the subgroups of all of the groups we achieve a new table:
The columns represent the groups and the rows represent the
subgroups (Table $II$).
We have already attained the initial requirements for
determining the cluster of a number, so therefore the
following lines describe the procedure:\\
If we want to determine that a number $X=(x_1, x_2, x_3)$ belongs to which cluster,\\\\
1- We build the table $II$\\
2- For finding the group: $r=x_2$ $\rightarrow$ we find r\\
3- For finding the subgroup:\\
\ 3-1- we compute $C(j, m).$mod$.p_2 \\ (C(j, m) = x_3)$\\
\ 3-2- $s(r, i) = C(j,m).$mod$.p_2$ $\rightarrow$ we find i ( we look up the table$II$)\\
4- We assign a relation to the subgroups of groups of all clusters:
General form of a relation of a number $N(j,m)$ that belongs to the subgroup i and group r and cluster m:\\\\
$re (i,j,m): (\ C(j,m) + i \times (p_3) + (m-1)\times p_2      \times p_3\ ).$mod$.p_1= A(j, m)$\\\\
$i\in (0,p_2-1) $\ \  number of subgroup\\
$j\in (0,p_3-1)$\\
$ m: $ the  number of the cluster $ 1\ to\ p_1$\\\\
\begin{table*}
\centering
\caption{the numbers that have $r_2$=r (group r) in cluster m}
\label{my-label}
\begin{tabular}{|c|c|c|c|}
\hline
\textbf {Num} & $\mathbf P_1$ & $\mathbf P_2$& $\mathbf P_3$\\
\hline
N(j,m) & A(j,m) & B(j,m) & C(j,m) \\
\hline
N(D(m)) & A(D(m)) & B(D(m)) & C(D(m)) \\
\hline
N(D(m)+1) & A(D(m)+1) & B(D(m)+1) & C(D(m)+1)\\
\hline
... & ... & ... & ... \\
\hline
N(D(m)+($P_3-1$)) & A(D(m)+($P_3-1$)) & B(D(m)+($P_3-1$)) & C(D(m)+($P_3-1$)) \\
\hline
\end{tabular}
\end{table*}%
4-1 After finding i the only unknown variable of the relation
equation is m $ (1 \ to \ p_1)$ that we solve the equation for m,
therefore we find the cluster that number is in it.\\\\
\begin{table}[b]
\begin{center}
\caption{The subgroups of group(r) for the numbers of
cluster m }
\label{my-label}
\begin{tabular}{ |c| } 
 \hline
 Subgroup(r)=Subgroups of group(r)\\  \hline
 $p_3$ \\  \hline
 S(r,i)  \\  \hline
 S(r,0)  \\  \hline
 S(r,1) \\  \hline
 ... \\  \hline
 S(r,$p_2$-1) \\  \hline
\end{tabular}
\end{center}
\end{table}\\
PROOF:\\\\
We use deductive reasoning to proof the general form of the
relations for every number in cluster(m) and group(r) , \\ \\      $ N(j,m) \ \ ( j \in (0,p_3-1), m \in  (1,p_1) ): $\\
\\ 1-First we proof that it is correct for \ $ j=0,i=0,m=0 $ :
\\ we should show that: \\ \\
$re (0, 0, 1): (\ C( 0 ,1) + 0 \times (p_3) + (1-1) \times (p_2 \times p_3)\ ).$mod$.p_1 = A(0,1) $ $\rightarrow$ $
(\ C(0,1) + 0 \times  (p_3) + (1-1) \times (p_2 \times  p_3)\ ).$mod$.p_1= r.$mod$.p_1 \ \ \ (1) $\\ \\
 $A (0,1) = r.$mod$.p_1 \ \ \ (2) $\\\\
From (1) and (2) we conclude that re(0,0,1) is correct.\\\\
2- Now we assume that relation re(i,j,m) is correct then we
should proof that it is correct for :\\ \\
$(C( j ,m)+i \times (p_3) + (m-1) \times p_2 \times p_3).$mod$.p_1= A(j,m)= (\ D(m) +(j\times p_2+r)\ ).$ mod$. p_1$\\\\
2-1 re(i, j+1, m)\\\\
We should proof that this relation is correct:\\\\
$re (i, j+1, m):$\\ $ (C( j+1 ,m) + i \times (p_3) + (m-1)\times (p_2 \times p_3)).$mod$.p_1 = A(j+1,m)$\\
$ (\ C( j+1 ,m) + i \times (p_3) + (m-1)\times(p_2\times p_3)\ ).$mod$.p_1= (\ (C( j ,m) + p_2) + i \times (p_3) + (m-1)\times(p_2 \times p_3)\ ).$mod$.p_1 \ (1)  $\\\\
since number of subgroup i haven’t changed 
\\\\$A(j+1,m) =(\ p_2+ D(m) +(j \times p_2+r)\ ).$ mod$.p_1 \ (2)$ \\\\
From (1) and (2) we conclude that the re(i, j+1, m) is correct,
because if $ ( a.$mod$.c= b.$mod$.c)$ then $(a+d).$mod$.c=(b+d).$mod$.c$  ( in this case c =$p_2$)\\\\
2-2 re(i+1, j+1, m) 
We should proof that this relation is correct:
$ re (i+1, j+1, m): (\ C(j+1,m) + (i+1) \times (p_3) + (m-1)\times (p_2 \times p_3)\ ).$mod$.p_1=A(j+1,m) $\\\\
$ (\ C( j+1 ,m) + (i+1) \times (p_3) + (m-1) \times (p_2 \times p_3)\ ).$mod$.p_1= (\ (( j+1) \times p_2+r). $mod$. p_3+ p_3 + i \times p_3 + (m-1) \times (p_2 \times p_3)).$mod$.p_1 $ \\\\, since     $ (\ p_2\ <\ p_3\ and\ i\rightarrow i+1) \rightarrow(\ (( j)\times p_2+r). $mod$. p_3 +p_2-p_3+ p_3 + i\times p_3 + (m-1)  \times (p_2 \times p_3)\ ).$mod$.p_1= ((( j)\times p_2+r). $mod$. p_3 +p_2 + i \times p_3 + (m-1) \times (p_2 \times p_3)).$mod$.p_1 \ (1) $\\\\
$ A (j+1, m) = (\ p_2 + D(m) + (j \times p_2+r)\ ). $mod$.p_1  \ (2) $\\\\
From (1), (2) we conclude that re (i+1, j+1, m) is correct.\\\\
2-3 re (i+1, j+1, m+1) We should proof that this relation is correct:\\
\\$ re(i+1,j+1,m+1): (C(j+1 ,m+1) + (i+1)\times (p_3) + ((m+1)-1) \times (p_2 \times p_3)).$mod$.p_1=A(j+1,m+1)\\
(C(j+1 ,m+1) + (i+1) \times (p_3) + ((m+1)-1) \times (p_2\times p_3)).$mod$.p_1,  $\\
since $ (p_2\ < \ p_3 \ $and$ \ i \rightarrow i+1) \ (1) $\\\\
$ (C(j,m)+ p_2 - p_3 + p_3 + i \times (p_3) + p_2 \times p_3+ (m-1) \times (p_2 \times p_3)).$mod$.p_1=( C(j ,m)+ p_2 + i \times  (p_3) + p_2 \times p_3 \ + (m-1) \times (p_2 \times p_3)).$mod$.p_1  $\\\\
$ A(j+1,m+1) = (D(m+1) +((j+1) \times p_2+r)).$ mod$. p_1= ( p_2 \times p_3 + p_2 + D(m) +(j \times p_2 + r)). $mod$. p_1\ (2) $\\\\
From (1) and (2) we conclude that re (i+1, j+1, m+1) is
correct because if $( a.$mod$.c= b.$mod$.c) $ then
\\$( (a+d).$mod$.c=(b+d).$mod$.c )$ ( in this case c = $p_2 \times p_3$ +$p_2$) \\\\
\begin{table}[]
\begin{center}
\caption{groups and subgroups }
\label{my-label}
\begin{tabular}{ |c|c|c|c| } 
 \hline
  S(0,0) & s(1,0)& ...& s($p_2$-1,0)  \\  \hline
  S(0,1) & s(1,1)& ...& s($p_2$-1,1)         \\  \hline
  ...    & ...   & ...& ...           \\  \hline
  S(0,$p_2$-1) & s(1,$p_2$-1)& ...&s($p_2$-1,$p_2$-1)   \\  \hline
 
\end{tabular}
\end{center}
\end{table}\\
Example 1:\\\\
For Modulo set (3 5 7) find out that the number (2 1 4) = 11 belongs to which cluster?\\\\
$P_1=3$ $\rightarrow$ we have three clusters.\\
First cluster: 0 to $(M/p_1)-1 = 0$ to 34\\
Second cluster: $(M/p_1)$ to $(2 \times M/p1)-1 = 35$ to 69\\
Third cluster: $2 \times M/p_1$ to $(3 \times M/p_1)-1 = 70$ to 104\\
We must show that the number (2 1 4) =11 belongs to the first cluster.\\\\
Solution:\\\\
1- $ r_2 = 1$ so the number belongs to the group 1. $\rightarrow$ r=1\\
2- We build the table III (to find the values of s (1, i)).\\
using (5) $\rightarrow$ S (1,i) : s (1,0) = 1, s (1, 1) = 4, s (1,2) = 2,
s (1, 3) = 0, s (1, 4) = 3\\
3- for finding the subgroup $r_3.$mod$.p_2=4.$mod$.5=4 $ $\rightarrow$ s (1, 1)= 4 $\rightarrow$ i=1\\
4- we want to find that which relation is correct for this
number therefore we could find out the cluster.\\
re (i, j, m): $(C(j,m).$mod$.p_1 + i \times(p_3.$mod$.p_1)+
(m-1) \times (p_2 \times p_3.$mod$.p_1)).$mod$.p_1 = A(j,m)$
$C(j,m)=r_3$ , $A(j,m)=r_1 $ $\rightarrow$ $r_3.$mod$.p_1 + i \times p_3.$mod$.p_1 + (m-1) \times (p_2 \times p_3.$mod$.p_1).$mod$.p_1= r_1
 Re(1,m):( 4.$mod$.3 + 1 \times 7.$mod$.3 + (m-1) \times (5 \times 7.$mod$.3)).$mod$.3= 2 $  $\rightarrow$  $ (2 \times m).$mod$.3=2$
  $\rightarrow$ \\
M:1 to $p_1$=3 $\rightarrow$ m=1  \\
The number is in the first cluster.\\\\
Example 2:\\\\ For Modulo set (3 5 7) find out that the
number (2 0 6)=20 belongs to which cluster?\\\\
Solution:\\\\
1- $r_2=0$ $\rightarrow$ r=0\\
2- S (0,i) : s (0,0) = 0 s (0, 1) = 3 s (0,2) = 1 s (0, 3) = 4 s (0, 4) = 2 \\
3- $r_3.$mod$.p_2=6.$mod$.5=1$ $\rightarrow$ s (0, 2) = 1 $\rightarrow$ i=2\\
4-$ Re(2,m):( 6.$mod$.3 + 2 \times 7.$mod$.3 +
(m-1)\times (5 \times 7.$mod$.3)).$mod$.3= 2 $ $\rightarrow$ $(2 \times m).$mod$.3=2 $ $\rightarrow$\\
M:1 \ to \ $p_1$=3 $\rightarrow$ m=1\\
The number is in the first cluster.\\\\
Example 3:\\\\ For Modulo set (3 5 7) find out that the
number (1 2 3)=52 belongs to which cluster?\\\\
Solution:\\\\
1- $r_2=2$ $\rightarrow$ $r=2$\\
2- S (2,i) : s (2,0) = 2 s (2, 1) = 0 s (2,2) = 3 s (2, 3) = 1 s (2, 4) = 4\\
3- $r_3.$mod$.p2=3.$mod$.5=3 $ $\rightarrow$ s (2, 2) = 3 $\rightarrow$ i=2\\
4- $Re(2,m):( 3.$mod$.3 + 2 \times 7.$mod$.3 + (m-1) \times (5 \times 7.$mod$.3)).$mod$.3=1 $ $\rightarrow$ $(2 \times m).$mod$.3=1 $ $\rightarrow$\\
$M:1 \ to \ p_1=3 $  $\rightarrow $  m=2\\
The number is in the second cluster.\\\\
Example 4:\\\\ For Modulo set (3 5 7) find out that the
number (0 3 0)=63 belongs to which cluster?\\\\
Solution:\\\\
1- $r_2=3$ $\rightarrow$ $ r=3$\\
2- S (3,i) : s (3,0) = 3 s (3, 1) = 1 s (3,2) = 4 s (3, 3) = 2 s (3, 4) = 0\\
3- $r_3.$mod$.p2=0.$mod$.5=0 $ $\rightarrow$ s (3, 4) = 0 $\rightarrow$ i=4\\
4- $Re(4,m):( 0.$mod$.3 + 4\times7.$mod$.3 +
(m-1)\times (5\times7.$mod$.3)).$mod$.3=0 $ $\rightarrow$ $ (2\times m+2).$mod$.3=0 $ $\rightarrow$\\
M:1 \ to \ $p_1$=3  $\rightarrow$  m=2\\
The number is in the second cluster.\\\\
Example 5:\\\\ For Modulo set (3 5 7) find out that the
number (0 1 5)=96 belongs to which cluster?\\
Solution:\\\\
1- $r_2=1$ $\rightarrow$ $ r=1$\\
2- S (1,i) : s (1,0) = 1 s (1, 1) = 4 s (1,2) = 2 s (1, 3) = 0 s (1, 4) = 3\\3-$ r_3.$mod$.p_2=5.$mod$.5=0 $ $\rightarrow$ $ s (1, 3) = 0$ $\rightarrow$ $ i=3 $\\
4-$ Re(3,m):( 5.$mod$.3 + 3 \times 7.$mod$.3 + (m-1)(5 \times 7.$mod$.3)).$mod$.3=0 $ $\rightarrow$ (2$\times m+3).$  mod.3=0 $\rightarrow$\\
$M:1\ to \ p_1=3  $ $\rightarrow$ m=3 \\
The number is in the third cluster.\\\\
\section{ Proposed Algorithm To Compare The Magnitude of Two Numbers Using Clustering Method}
In order to compare the magnitude of two numbers, X and Y,
the following algorithm is proposed.\\
1. Consider modulo set P=$(p_1, p_2, p_3)$.\\
2. Input X= $( x_1, x_2, x_3)$, where $x_i= X.$mod$.p_i$ for i=1 to 3 \\
3. Input Y= $( y_1, y_2, y_3)$, where$ y_i= Y.$mod$.p_i $ for i=1 to 3 \\
4. Find Z=X-Y in the given modulo set $Z= ( z_1, z_2, z_3)$, where\ $ z_i= (x_i-y_i).$mod$.p_i $for i=1 to 3\\
5. Find cluster of X, CL(X), using section.2\\
6. Find cluster of Y, CL(Y), using section.2\\
7. Find cluster of Z, CL(Z), using section.2\\
8. If Z=0 then X=Y else\\
If CL(X) $>$ CL(Y) then X $>$ Y End if.\\
If CL(X) $<$ CL(Y) then X $<$ Y End if.\\
If CL(X) = CL(Y) .and. CL(Z) = 1 then X$ >$ Y End if.\\
If CL(X) = CL(Y) .and. CL(Z) = $p_1$ then X$ <$ Y End if.\\
End if.\\\\
Fig.1. Algorithm to Compare Two Numbers Using Clustering
Method\\\\
Example 1:\\\\  For modulo set (3 5 7) find out that which one of these two numbers have a greater magnitude? \\
X= (0, 1, 5)=96\\
Y= (2, 1, 4)=11\\\\
Solution:\\\\
We use algorithm in fig.1\\
1- We find Z $\rightarrow$ Z=(1, 0, 1) $\neq$ 0\\
2- We find CL(X) $\rightarrow$ CL(X) =3 (from example 5 section 2)\\
Note: CL(X) = 1 means that X belongs to the first cluster.\\
CL(X) = 2 means that X belongs to the second cluster.\\
CL(X) = 3 means that X belongs to the third cluster.\\
3- We find CL(Y) $\rightarrow$ CL(Y) = 1 (from example 1 section 2)\\
4- We compare CL(X) and CL(Y) :\\
4-1- CL(X) =3\\
4-2- CL(Y) =1\\
From 4-1 and 4-2 $\rightarrow $CL(X)$ >$ CL(Y)\\
5- Z $\neq$ 0, CL(X)$ >$ CL(Y) $\rightarrow$ X$> $Y (from Fig 1)\\\\
Example 2:\\\\ For modulo set (3 5 7) find out that which one of these two numbers have a greater magnitude?\\
X= (2, 1, 4) =11
Y= (0, 1, 5) =52\\\\
Solution: \\\\
1- Z$\neq$0\\
2- CL(X) = 1 (from example1 section 2)\\
3- CL(Y) = 2 (from example 3 section 2)\\
4- CL(Y) $>$ CL(X) $\rightarrow$ Y $>$ X\\\\
Example 3: \\\\For modulo set (3 5 7) find out that which one of these two numbers have a greater magnitude?\\
X= (0, 3, 0) =63\\
Y= (0, 1, 5) =52\\\\
Solution :\\\\
1- Z $\neq$  0\\
2- CL(X) = 2 (from example 4 section 2)\\
3- CL(Y) = 2 (from example 3 section 2)\\
4- Z = X-Y = 11 = (2 1 4)\\
5- CL(Z) = 1 (from example 1 section 2) $\rightarrow$ X $>$ Y\\
\section{ Boolean Circuit}
The Boolean circuit for cluster finding algorithm for the moduli set (2,3,5) has been presented below the input is any number $N=(N_1,N_2,N_3)$ which belongs to this cluster.
\\\\
$N_1=(N_11,N_12)$ \\\\
$N_2=(N_21,N_22)$\\\\
$N_3=(N_13,N_23,N_33)$\\\\
$\{N_11,N_12,N_21,N_22,N_13,N_23,N_33\} \in \{0,1\}$\\\\
if $OUT=0$ $\rightarrow$ the number belongs to the first cluster. \\
if $OUT=1$ $\rightarrow$ the number belongs to the second cluster.
\\
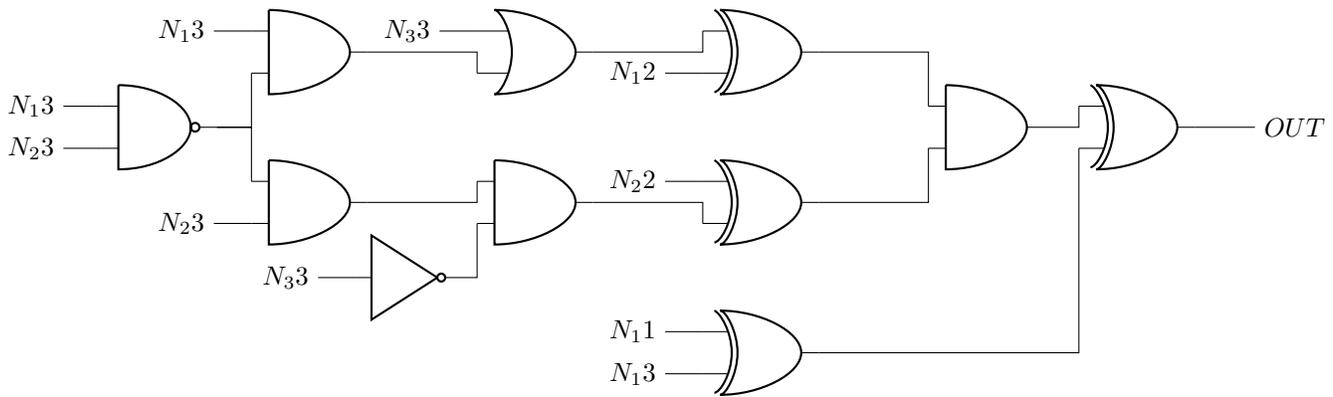
\begin{circuitikz}\draw
(8,0) node[xor port] (xor1) {}
(8,2) node[xor port] (xor2) {}
(13,1) node[xor port] (xor3) {}
(11,1) node[and port] (myand4) {}
(8,-2) node[xor port] (xor4) {}
(2,2) node[and port] (myand1) {}
(2,0) node[and port] (myand2) {}
(5,0) node[and port] (myand3) {}
(5,2) node[or port] (myor) {}
(0,1) node[nand port] (nand1) {}
(2.7,-1) node[not port] (not1) {}
(myand4.out) -| (xor3.in 1) 
(xor4.out) -| (xor3.in 2) 
(xor1.out) -| (myand4.in 2)
(xor2.out) -| (myand4.in 1)
(myor.out) -| (xor2.in 1)
(myand3.out) -| (xor1.in 2)
(not1.out) -| (myand3.in 2)
(myand1.out) -| (myor.in 2)
(myand2.out) -| (myand3.in 1)
(nand1.out) -| (myand1.in 2)
(nand1.out) -| (myand2.in 1);
\draw (xor4.in 2) -- ++(left:5mm) node[left] (A) {$N_13$};
\draw (xor4.in 1) -- ++(left:5mm) node[left] (A2) {$N_11$};
\draw (nand1.in 1) -- ++(left:5mm) node[left] (A) {$N_13$};
\draw (nand1.in 2) -- ++(left:5mm) node[left] (B) {$N_23$};
\draw (not1.in) -- ++(left:5mm) node[left] (C) {$N_33$};
\draw (myand1.in 1) -- ++(left:5mm) node[left] (A) {$N_13$};
\draw (myand2.in 2) -- ++(left:5mm) node[left] (B) {$N_23$};
\draw (myor.in 1) -- ++(left:5mm) node[left] (C) {$N_33$};
\draw (xor2.in 2) -- ++(left:5mm) node[left] (A1) {$N_12$};
\draw (xor1.in 1) -- ++(left:5mm) node[left] (B1) {$N_22$};
\draw (xor3.out) -- ++(right:8mm) node[right] (B1) {$OUT$};
\end{circuitikz}\\
\section{Conclusion}
Comparison of two numbers, like division of two numbers, in
RNS systems is considered a difficult task. In this paper, a
new algorithm to compare the magnitude of two RNS
numbers, using a clustering method, was proposed. The
modulo set of RNS system was considered to be {t,p,q}. In
the proposed algorithm, first, a clustering process was
introduced in order to assign a cluster to each inputted
number. To compare the magnitude of two numbers, the
clusters of these numbers and their difference were obtained.
Then by comparing those clusters, the relative magnitude of
two numbers was determined. All of those processes were
performed in RNS system without converting numbers to the
binary system.


\begin{thebibliography}{6}
\bibitem{gdimuroa}G. Dimauro, S. Impedovo, and G. Pirlo " A new technique for fast number comparison in residue number system" \emph{IEEE Trans. on Comp}, pp. 608-612, 1993.
\bibitem{Miller}D. D. Miller, R. E. Altschul, J. R. King, and J. N. Polky " Analysis of the residue class core function of Akushskii, Burcev and Pak. In Residue number System Arithmetic: Modern Applications in Digital Signal Processing" \emph{IEEE Press, New York}, pp. 390-401, 1986.
\bibitem{Szabo}N. S. Szabo and R. I. Tanaka "  Residue Arithmetic and its Applications to Computer Technology" \emph{McGraw-Hill}, 1967.
\bibitem{wang}Y. Wang "  New chinese remainder theorems In Proc. Of Thirty-Second Asilomar Conference on Signals, Systems and Computers"\emph{McGraw-Hill},volume 1, pp. 165-171 , 1998.
\bibitem{wang,song}Y.Wang, X. Song, and M. Aboulhamid "   A new algorithm for RNS magnitude comparison based on new chinese remainder theorem II. In Proc.of the Ninth Great Lakes Symposium on VLSI", pp. 362-365, 1999.
\bibitem{taylor}F. J. Taylor,"Residue arithmetic: A tutorial with examples,"\emph{IEEE Comput. Mag.},vol. 17, no. 5, pp. 50-62, May 1984.
\end{thebibliography}
\end{document}